\documentclass[11pt]{article}
 
\usepackage[final]{acl}
 \usepackage[most]{tcolorbox}
\usepackage{times}
\usepackage{latexsym}
\usepackage{booktabs,multirow,longtable}
\usepackage[T1]{fontenc}
 
\usepackage[utf8]{inputenc}
 
\usepackage{microtype}
\usepackage{graphicx}
\usepackage{subcaption}
\usepackage{inconsolata}
 
\usepackage{graphicx}
 \usepackage{makecell}
\usepackage{multirow}
\usepackage{colortbl}
\usepackage{xcolor}
\usepackage{bm}
\usepackage{enumitem}
\usepackage{mathtools}
\usepackage{booktabs}
\usepackage{comment}
\usepackage{tcolorbox}
\tcbuselibrary{breakable, skins}
\usepackage{enumitem}
\usepackage{float}
\usepackage{algorithm}
\usepackage{algpseudocode}
\usepackage{amssymb}
\usepackage[normalem]{ulem}

\title{ADORE: Iterative Query Expansion with Retrieval-Grounded Relevance Feedback}

\author{
  \textbf{Amin Bigdeli\textsuperscript{1}},
  \textbf{Negar Arabzadeh\textsuperscript{2}},
  \textbf{Radin Hamidi Rad\textsuperscript{3}},
  \\
  \textbf{Sajad Ebrahimi\textsuperscript{3}},
  \textbf{Charles L. A. Clarke\textsuperscript{1}},
  \textbf{Ebrahim Bagheri\textsuperscript{3}}
  \\
  \textsuperscript{1}University of Waterloo
  \\
  \textsuperscript{2}University of California, Berkeley
  \\
  \textsuperscript{3}University of Toronto
  \\
  {\small\textbf{Correspondence:}
  \href{mailto:abigdeli@uwaterloo.ca}{abigdeli@uwaterloo.ca}}
  \\
  \texttt{\{abigdeli,charles.clarke\}@uwaterloo.ca}
  \\
  \texttt{negara@berkeley.edu}
  \\
  \texttt{\{radin.rad,s.ebrahimi,ebrahim.bagheri\}@utoronto.ca}
}

\begin{document}

\maketitle

\begin{abstract}
LLM-based query expansion improves retrieval by enriching the query with additional context. Yet most methods remain generation-driven, producing plausible expansions without checking how the target corpus responds. This can introduce retrieval drift, amplify misleading vocabulary, or miss terms that distinguish relevant from non-relevant documents. We argue that effective query expansion requires retrieval-grounded feedback, not just single-pass generation or unverified iteration. We introduce \texttt{ADORE} (\textbf{AD}apt, \textbf{O}bserve, \textbf{R}elevance \textbf{E}valuate), an iterative framework that turns retrieval outcomes into feedback for subsequent query expansions. At each round, an LLM generates pseudo-passages, a retriever exposes the corpus response, and a relevance assessor evaluates retrieved documents against the original query. These judgments identify what to reinforce, what remains undercovered, and what to suppress. Across TREC Deep Learning, BEIR, and BRIGHT, \texttt{ADORE} consistently outperforms strong query expansion baselines with notable improvements across nearly all evaluation settings. It improves average nDCG@10 on BEIR by 24.5\% over BM25 and by 3.6\% over the strongest query expansion baseline. On the reasoning-intensive BRIGHT benchmark, it more than doubles BM25 performance from 0.170 to 0.379 and surpasses the best query expansion baseline by 9.2\%. Our code and data are publicly available at: \url{https://github.com/aminbigdeli/ADORE}.
\end{abstract}
 
\section{Introduction}
\label{sec:intro}
Query reformulation has long been an effective way to improve retrieval, not only by reducing vocabulary mismatch, but also by clarifying underspecified queries, surfacing missing facets, and adapting user queries to the language of the target corpus~\citep{rocchio1971relevance,lavrenko2001relevance,abdul2004umass,bhogal2007review,qiu1993concept}. Large language models (LLMs) have made this strategy more flexible by generating synthetic expansion text that broadens query coverage and introduces potentially useful terminology.

Recent work has explored a range of LLM-based reformulation strategies, including pseudo-document generation~\citep{wang2023query2doc}, instruction-based reformulation~\citep{wang2023genqr,dhole2024genqrensemble,reformer_2026,exploringresearch}, hypothetical answer generation~\citep{gao2022precise}, sub-question decomposition~\citep{qaexpand}, corpus-grounded reformulation~\citep{lamer,csqe}, and aggregation across multiple generations~\citep{mugi}. Together, these methods have shown that LLM-generated expansion text can substantially improve retrieval effectiveness across standard benchmarks.

Despite these advances, existing LLM-based reformulation methods remain limited in three ways.
\textbf{Lack of retrieval-outcome awareness.} Most methods generate an expansion and pass it to the retriever, but never inspect the ranking it induces. They therefore cannot tell which terms improved retrieval, which were ineffective, and which caused retrieval drift. This creates a gap between semantic plausibility and retrieval utility, e.g., an expansion can sound useful while still retrieving off-topic documents or missing domain-specific terminology.

\noindent \textbf{Shallow use of corpus signals.} Existing corpus-aware methods retrieve documents, but do not turn them into explicit feedback. LameR uses retrieved passages as in-context examples~\citep{lamer}, CSQE extracts salient sentences from initial results~\citep{csqe}, and ThinkQE feeds retrieved documents back into the generator across rounds~\citep{thinkqe}. In all cases, retrieved documents mainly serve as generation context, not as evidence for diagnosing retrieval success, missing query facets, or false-positive signals.

\noindent \textbf{Unverified iteration.} Iterative methods like ThinkQE~\cite{thinkqe} assume that feeding retrieved documents back into the generator will progressively improve the reformulation, but they do not verify that each round improves retrieval. Without explicit retrieval-grounded assessment, later rounds may reinforce misleading terms, drift toward false-positive corpus regions, or repeat earlier retrieval errors rather than correct them.

Together, these limitations suggest a different view of query reformulation. The goal is not to generate expansion text that merely sounds relevant, but to generate text that improves retrieval on the target corpus. This requires \emph{retrieval-grounded feedback}, i.e., observing how the corpus responds, assessing the retrieved evidence, and using that assessment to guide the next reformulation.
Under this view, query reformulation becomes an iterative retrieval refinement process rather than a text generation problem. Each round should identify signals that retrieve relevant evidence, partially match the information need, or cause retrieval drift, allowing the system to reinforce useful corpus-specific vocabulary while suppressing misleading terms.

To address this gap, we introduce \texttt{ADORE} (\textbf{AD}apt, \textbf{O}bserve, \textbf{R}elevance \textbf{E}valuate), an iterative query reformulation framework that turns retrieval outcomes into feedback for the next reformulation. \texttt{ADORE} operates in three steps. First, it generates pseudo-passages conditioned on the original query and feedback from previous rounds. Second, it retrieves documents from the target corpus to expose how the corpus responds to the current reformulation. Third, it assesses the newly retrieved documents against the original query and converts them into graded feedback that identifies what to reinforce, what remains undercovered, and what to suppress.
This feedback loop makes reformulation retrieval-grounded rather than generation-only. \texttt{ADORE} refines expansions based on observed corpus behavior, while anchoring all relevance assessments to the original query to prevent retrieval drift across rounds. 
\texttt{ADORE} differs from prior work by separating generation from evaluation. Rather than feeding raw retrieved passages back to the generator and hoping it infers what matters, \texttt{ADORE} explicitly evaluates retrieval outcomes and turns them into structured feedback. This makes later rounds corrective rather than merely iterative.

We evaluate \texttt{ADORE} on TREC Deep Learning (DL), BEIR, and BRIGHT, spanning passage retrieval, cross-domain retrieval, and reasoning-intensive retrieval. \texttt{ADORE} consistently outperforms strong query reformulation baselines, achieving notable gains across nearly all evaluation settings. These results show that iterative reformulation is most effective when each round is guided by explicit retrieval-grounded feedback, rather than by single-pass generation or unstructured reuse of retrieved documents.
Our contributions can be summarized as follows:

\begin{itemize}[leftmargin=*,nosep]

\item We introduce \texttt{ADORE}, a retrieval-grounded query reformulation framework that separates generation from evaluation by iteratively generating expansions, observing corpus responses, and using relevance feedback to guide the next round.

\item We instantiate \texttt{ADORE} with an adaptive refinement loop that keeps relevance assessment anchored to the original query and stops when retrieval quality or coverage saturates.

\item Through extensive experiments on TREC DL, BEIR, and BRIGHT, we demonstrate that \texttt{ADORE} consistently outperforms strong reformulation baselines, improving average nDCG@10 by 24.5\% over BM25 and 3.6\% over the strongest prior reformulation method on BEIR, and raising it from 0.170 to 0.379 over BM25 on BRIGHT, a 9.2\% gain over the best reformulation baseline, with notable improvements across nearly all evaluation settings. 
 
\end{itemize}

\section{Related Work}
\label{sec:related_work}

\paragraph{LLM-driven query reformulation.}
Recent methods leverage LLMs for query expansion and can be grouped into three categories. Keyword-based methods generate expansion terms to enrich the query, including GenQR~\citep{wang2023genqr}, GenQREnsemble~\citep{dhole2024genqrensemble}, and Query2E~\citep{exploringresearch}. Response-based methods generate pseudo-passages or answers that are appended to the query, including Query2Doc~\citep{wang2023query2doc}, MUGI~\citep{mugi}, and QA-Expand~\citep{qaexpand}. Corpus-aware methods further condition the expansion on retrieved documents, including LameR~\citep{lamer}, CSQE~\citep{csqe}, and ThinkQE~\citep{thinkqe}. Across these families, methods either operate zero-shot, relying solely on the model's parametric knowledge, or draw on corpus signals, with some applying only an implicit notion of relevance and others iterating for a fixed number of rounds without verifying when to stop. Zero-shot methods like MUGI expand without observing how the corpus responds, while corpus-aware methods like LameR and CSQE use retrieved documents as generation context but neither check whether the reformulation improves retrieval nor revisit the corpus. ThinkQE iterates over multiple rounds, yet it feeds raw passages back to the generator and assumes most are irrelevant, without assessing which documents actually help or mislead. \texttt{ADORE} addresses these limitations of prior work by iteratively refining the query expansion, observing the corpus each round, converting every retrieved document into a point-wise graded relevance signal that reinforces relevant and suppresses misleading evidence, and stopping through adaptive termination (Table~\ref{tab:method_comparison}).

\begin{table}[t]
\centering
\small
\setlength{\tabcolsep}{5pt}
\providecommand{\fullmark}{\textcolor{green!40!black}{$\checkmark$}}
\providecommand{\partmark}{\textcolor{orange!70!black}{$\sim$}}
\providecommand{\nomark}{\textcolor{red!60!black}{$\times$}}
\scalebox{0.86}{
\begin{tabular}{l cccc}
\toprule
\textbf{Method} & \makecell{\textbf{Corpus}\\\textbf{Signal}} & \textbf{Iterative} & \makecell{\textbf{Reinforce}\\\textbf{\& Suppress}} & \makecell{\textbf{Adaptive}\\\textbf{Stop}} \\
\midrule
\texttt{GenQR}          & \nomark   & \nomark   & \nomark   & \nomark \\
\texttt{GenQREnsemble}  & \nomark   & \nomark   & \nomark   & \nomark \\
\texttt{QA-Expand}      & \nomark   & \nomark   & \partmark & \nomark \\
\texttt{Query2Doc}      & \nomark   & \nomark   & \nomark   & \nomark \\
\texttt{LameR}          & \fullmark & \nomark   & \nomark   & \nomark \\
\texttt{MUGI}           & \nomark   & \nomark   & \nomark   & \nomark \\
\texttt{CSQE}           & \fullmark & \nomark   & \partmark & \nomark \\
\texttt{ThinkQE}        & \fullmark & \fullmark & \nomark   & \nomark \\
\midrule
\textbf{\texttt{ADORE}} & \fullmark & \fullmark & \fullmark & \fullmark \\
\bottomrule
\end{tabular}
}
\caption{Comparison of \texttt{ADORE} against query reformulation methods. \fullmark~fully supported, \partmark~partially or implicitly supported, \nomark~not supported.}
\vspace{-1em}
\label{tab:method_comparison}
\end{table}

\paragraph{LLM-as-judge for retrieval.}
The use of LLMs as relevance assessors has received growing attention \cite{Arabzadeh_2025,rahmani2024llmjudge,Shivani_ICTIR,farzi2025does}. \citet{thomas2024large} and the UMBRELA framework~\citep{upadhyay2024umbrela} demonstrated strong agreement between LLM and human relevance judgments on TREC benchmarks, with UMBRELA showing that open-weight models produce competitive graded assessments. \citet{macavaney2023one} found that even minimal supervision yields useful relevance labels. This agreement generalizes across different LLMs~\citep{farzi2025does}, indicating that LLM relevance judgments offer a reliable signal for guiding \texttt{ADORE}'s expansion. While LLM-based relevance assessment has largely been used for offline evaluation, our work repurposes graded relevance assessment as an online feedback signal within an iterative retrieval loop, using the assessor's structured judgments to steer subsequent expansion generation.

\section{Methodology}
\label{sec:method}

\begin{figure*}[t]
    \centering

    \includegraphics[width=1.07\textwidth,
            trim=4cm 4cm 0cm 4.3cm,
        clip
        ]{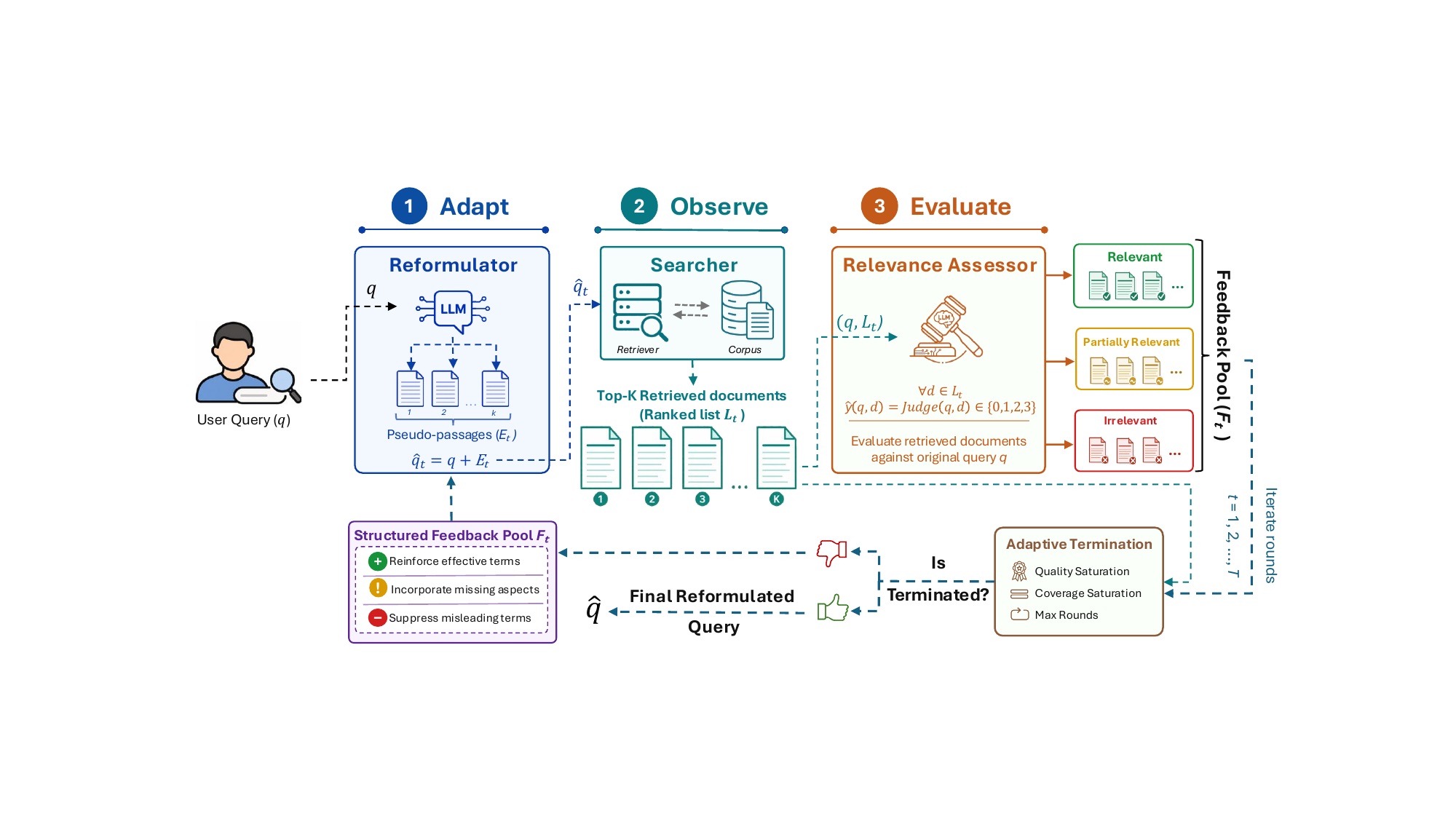}
\caption{Overview of \texttt{ADORE}. Each round consists of three stages: \textbf{Adapt} generates corpus-calibrated pseudo-passages conditioned on structured feedback from the previous round, \textbf{Observe} retrieves documents from the target corpus, and \textbf{Evaluate} assesses the results to produce structured feedback for the next iteration.}

\label{fig:adore_workflow}

\end{figure*}

 \subsection{Problem Formulation}
\label{sec:problem}
 
We formulate query reformulation as a retrieval-grounded optimization problem. Given an input query $q$, a corpus $\mathcal{C}$ of documents $d \in \mathcal{C}$, a retrieval function $\mathcal{R}$, and a ranking evaluation metric $\mathcal{M}$, the objective is to find the reformulated query $\hat{q}^\star$ that, among all candidate reformulations $\hat{q}$, maximizes retrieval quality over the induced ranking:

\begin{equation}
\hat{q}^\star = \arg\max_{\hat{q}} \; \mathcal{M}\!\left(\mathcal{R}(\hat{q};\, \mathcal{C}),\; \bm{y}^\star(q)\right)
\label{eq:objective}
\end{equation}

where $\bm{y}^\star(q) = \{y^\star(q, d)\}$ denotes the true graded relevance labels for $q$. In practice, these relevance labels are unavailable at inference time, making it impossible to directly optimize the reformulation against the true retrieval objective.
 
Existing methods approximate this objective mainly through generation. Given $q$, an LLM generates an expansion or pseudo-passage, $\hat{q}=g_\theta(q)$, and the reformulated query is then issued to the retriever. The retrieval outcome is typically accepted as-is. Even corpus-aware methods that include retrieved passages in the prompt often treat them as context for generation, rather than as evidence whose retrieval utility should be evaluated. As a result, existing methods lack an explicit mechanism for diagnosing which generated signals improve retrieval, which cause drift, and which aspects of the information need remain uncovered.

 \subsection{\texttt{ADORE} Query Reformulation}
\label{sec:framework}
 
Motivated by this gap, \texttt{ADORE} treats query reformulation as iterative retrieval refinement, built on three principles. First, reformulation should be retrieval-grounded, i.e., expansion should be judged by the ranking it induces, not only by how plausible it sounds. Second, corpus signals should serve as feedback, not just generation context, i.e., retrieved documents should reveal what works, what is missing, and what causes false positives. Third, iteration should be goal-oriented, i.e., each round uses retrieval behavior to correct the previous one while remaining anchored to the original query.

\texttt{ADORE} implements these principles through three stages. First, a reformulator generates pseudo-passages conditioned on the original query and prior feedback if available. Second, a retriever issues the reformulated query and exposes the induced ranking. Third, a relevance assessor evaluates newly retrieved documents against the original query and converts them into structured feedback for the next round. Thus, retrieved documents serve not as raw context but as evidence of what to reinforce, what is missing, and what to suppress.

Formally, \texttt{ADORE} runs for up to $T$ rounds while constructing at each round a feedback pool $\mathcal{F}_t$ of graded relevance assessments. At each round, the reformulator generates $n$ pseudo-passages $E_t = \{e_t^{(1)}, \ldots, e_t^{(n)}\}$ conditioned on the original query $q$ and the previous round's feedback pool $\mathcal{F}_{t-1}$:

\begin{equation}
E_t = g_\theta(q,\, \mathcal{F}_{t-1})
\label{eq:conditioned}
\end{equation}

At each round, the top retrieved documents are assessed against the original query to form $\mathcal{F}_t$, exposing useful signals, missing facets, and sources of retrieval drift to the next reformulation. The following subsections describe these stages in detail. Figure~\ref{fig:adore_workflow} provides an overview of the framework.
 
\subsubsection{Adapt: Feedback-Conditioned Reformulation}
\label{sec:reformulator}
 At each round $t$, the Adapt stage generates a set of $n$ pseudo-passages $E_t$ through an LLM reformulator $g_\theta$. In the first round, no corpus feedback is available, so the reformulator operates in a zero-shot setting $
E_1 = \{e_1^{(i)} = g_\theta(q)\}_{i=1}^{n}$.
These passages provide the initial expansion from the query alone based on the LLM's parametric knowledge, a starting point for subsequent corpus-grounded refinement. The prompt is given in Appendix~\ref{sec:prompt_zeroshot}.

In subsequent rounds ($t \geq 2$), the reformulator conditions on the feedback pool $\mathcal{F}_{t-1}$ produced by the preceding evaluation stage (Equation~\ref{eq:conditioned}; full prompt in Appendix~\ref{sec:prompt_iter}). This pool partitions the previous round's retrieved evidence into graded relevance tiers, each providing a distinct adaptation signal. Higher-tier documents indicate patterns to reinforce, while lower-tier documents reveal under-represented aspects of the information need or false-positive attractors to suppress.
 
\subsubsection{Observe: Retrieval Outcome}
\label{sec:searcher}
 
The Observe stage exposes how the corpus actually responds to the current expansion. At each round $t$, the pseudo-passages $E_t$ generated in the Adapt stage are combined with the original query through a query construction function:
$\hat{q}_t = f(q, E_t)
\label{eq:query}
$
where $f$ balances the contribution of the original query and the current expansion text. By grounding each retrieval on only the latest round's pseudo-passages, the framework ensures that each round's observation reflects the reformulator's most recent adaptation to the feedback signal, preventing stale or superseded expansions from diluting the query. The constructed query is then issued to the retriever, yielding a ranked list:

\begin{equation}
L_t = \mathcal{R}(\hat{q}_t;\, \mathcal{C}) = \langle d_{t,1}, d_{t,2}, \ldots, d_{t,N} \rangle
\label{eq:retrieval}
\end{equation}

This ranked list constitutes the system's observation of how the corpus responds to the current round's expansion. By examining $L_t$ at each round, the framework can detect whether the latest pseudo-passage is surfacing relevant documents, drifting toward off-topic regions, or failing to capture important facets of the query. The specific instantiation of $f$ is described in Section~\ref{sec:setup}.
 
\subsubsection{Evaluate: Graded Relevance Assessment}
\label{sec:assessor}

\begin{table*}[!t]
\centering
\scalebox{0.75}{%
\begin{tabular}{l|l|ccc||ccccc|c}
\toprule
\multirow{2}{*}{\textbf{Category}} & \multirow{2}{*}{\textbf{Method}} 
& \multicolumn{3}{c||}{\textbf{TREC Deep Learning}} 
& \multicolumn{6}{c}{\textbf{BEIR Benchmark}} \\
& & \textbf{DL19} & \textbf{DL20} & \textbf{DLH} 
& \textbf{SciFact} & \textbf{COVID} & \textbf{FiQA} & \textbf{DBPed} & \textbf{NEWS} & \textbf{Avg.} \\
\midrule

\multirow{2}{*}{\makecell[l]{\textit{Sparse}}}
& \texttt{BM25} & 0.506 & 0.480 & 0.285 & 0.679 & 0.595 & 0.236 & 0.318 & 0.395 & 0.445 \\
& \texttt{BM25 + RM3} & 0.515 & 0.492 & 0.264 & 0.646 & 0.593 & 0.192 & 0.308 & 0.426 & 0.433 \\

\midrule
\multirow{4}{*}{\makecell[l]{\textit{Dense}}}
& \texttt{DPR} & 0.622 & 0.653 & -- & 0.318 & 0.332 & 0.295 & 0.263 & 0.161 & 0.274 \\
& \texttt{ANCE} & 0.645 & 0.646 & 0.334 & 0.570 & 0.654 & 0.300 & 0.281 & 0.382 & 0.437 \\
& \texttt{Contriever-FT} & 0.621 & 0.632 & 0.396 & 0.677 & 0.596 & 0.329 & 0.413 & 0.428 & 0.489 \\
& \texttt{BGE-base-en-v1.5} & 0.702 & 0.677 & 0.379 & 0.741 & 0.780 & 0.407 & 0.407 & 0.442 & 0.555 \\

\midrule
\multirow{11}{*}{\textit{\makecell[l]{Query\\Expansions\\w/ \texttt{BM25}}}}
& \texttt{GenQR} & 0.548 & 0.537 & 0.292 & 0.726 & 0.687 & 0.230 & 0.344 & 0.465 & 0.490 \\
& \texttt{GenQREnsemble} & 0.559 & 0.553 & 0.270 & 0.725 & \underline{0.753} & 0.239 & 0.360 & 0.486 & 0.513 \\
& \texttt{Query2E} & 0.594 & 0.576 & 0.345 & 0.709 & 0.715 & 0.269 & 0.378 & 0.463 & 0.507 \\
& \texttt{QA-Expand} & 0.683 & 0.642 & 0.302 & 0.706 & 0.707 & 0.264 & 0.370 & 0.450 & 0.499 \\
& \texttt{Query2Doc (CoT)} & 0.653 & 0.624 & 0.329 & 0.714 & 0.728 & 0.258 & 0.393 & 0.466 & 0.512 \\
& \texttt{Query2Doc (FS)} & 0.690 & 0.675 & 0.356 & 0.712 & 0.708 & 0.268 & 0.401 & 0.480 & 0.514 \\
& \texttt{Query2Doc (ZS)} & 0.687 & 0.663 & 0.350 & 0.720 & 0.743 & 0.260 & 0.406 & 0.498 & 0.525 \\
& \texttt{LameR} & 0.637 & 0.653 & 0.356 & 0.725 & 0.702 & 0.262 & 0.399 & 0.480 & 0.514 \\
& \texttt{MUGI} & 0.695 & 0.658 & 0.365 & \underline{0.735} & 0.714 & 0.264 & \underline{0.410} & \underline{0.516} & 0.528 \\
& \texttt{CSQE} & 0.690 & 0.655 & 0.366 & 0.721 & 0.699 & 0.247 & 0.390 & 0.479 & 0.507 \\
& \texttt{ThinkQE} & \underline{0.697} & \underline{0.683} & \underline{0.372} & \underline{0.735} & 0.733 & \underline{0.273} & \textbf{0.420} & 0.512 & \underline{0.535} \\[0.3em]

\cline{2-11}
\cline{2-11} \\[-0.87em]

 & \textbf{\texttt{ADORE}} & \textbf{0.713} & \textbf{0.712} & \textbf{0.383} & \textbf{0.755} & \textbf{0.774} & \textbf{0.315} & 0.407 & \textbf{0.520} & \textbf{0.554} \\

\bottomrule
\end{tabular}%
}
\caption{Results on TREC Deep Learning and BEIR benchmarks (nDCG@10). All query reformulation baselines use GPT-4.1 as the backbone language model. Best scores in \textbf{bold}, second-best \underline{underlined}.}
\vspace{-0.5em}
\label{tab:comprehensive_results}
\end{table*}
 
The Evaluate stage analyzes the current retrieval outcome and converts it into structured feedback for subsequent reformulation. For each newly retrieved document $d$ that has not been assessed before, a relevance assessor assigns a graded label:

\begin{equation}
\hat{y}(q, d) = \mathcal{J}_\phi(q, d) \in \{0, 1, 2, 3\},
\label{eq:judge}
\end{equation}
where $\mathcal{J}_\phi$ denotes the assessor, including the underlying LLM, prompt, and decoding configuration. Relevance is always assessed with respect to the original query $q$, rather than the reformulated query $\hat{q}_t$. This keeps feedback anchored to the user's information need and prevents reformulation drift from propagating across rounds.
 
Let $\mathcal{S}_{t-1} = \bigcup_{i=1}^{t-1} D_i^{\mathrm{new}}$ denote the documents assessed before round $t$. At each iteration, the assessor considers the top $K$ documents in $L_t$ that have not yet received a relevance label:

\begin{equation}
D_t^{\mathrm{new}} = \{d \in L_t \setminus \mathcal{S}_{t-1} \mid \mathrm{rank}_{L_t}(d) \leq K\}
\label{eq:newdocs}
\end{equation}

The feedback pool is rebuilt at each round from the top $K$ documents returned in the current round, paired with their graded labels:

\begin{equation}
\mathcal{F}_t = \{(d, \hat{y}(q,d)) : d \in L_t,\ \mathrm{rank}_{L_t}(d) \leq K\}
\label{eq:feedback_pool}
\end{equation}

Since relevance is evaluated against the original query $q$, labels are freshly assigned only to newly surfaced documents and reused otherwise, avoiding reassessment and focusing computation on new evidence. Grounding $\mathcal{F}_t$ in the documents retrieved in the current round keeps the feedback aligned with the query's most refined expansion, providing the reformulator with the most informative signal of how that expansion behaves on the corpus.

\subsubsection{Iterative Refinement and Adaptive Termination}
\label{sec:loop}
 
The adapt--observe--evaluate cycle repeats across rounds, with each iteration refining the expansion using an updated feedback pool. Instead of running for a fixed number of rounds, \texttt{ADORE} uses adaptive termination to allocate more computation to harder queries. The loop stops when one of three conditions is met: (i)~\emph{quality saturation}, where all assessed documents in the current round receive the maximum relevance score; (ii)~\emph{coverage saturation}, where the number of newly retrieved documents $|D_t^{\mathrm{new}}|$ falls below a threshold $\tau$ for two consecutive rounds; or (iii)~\emph{budget exhaustion}, where the round counter reaches the maximum $T$.
 
After termination at round~$T'$, final retrieval is performed using $\hat{q}_{T'}$. Since each round refines the expansion using the previous round's feedback, the final query uses the last, most refined expansion.

\section{Experimental Setup}
\label{sec:setup}

\begin{table*}[t]
\centering

\scalebox{0.76}{%
\begin{tabular}{l|l|c|ccccccc|c}
\toprule
\multirow{2}{*}{\textbf{Category}} & \multirow{2}{*}{\textbf{Method}} & \multirow{2}{*}{\textbf{Training}} & \multicolumn{8}{c}{\textbf{BRIGHT Benchmark}} \\
& & & \textbf{Bio.} & \textbf{Earth.} & \textbf{Econ.} & \textbf{Psy.} & \textbf{Rob.} & \textbf{Stack.} & \textbf{Sus.} & \textbf{Avg.} \\
\midrule

\multirow{2}{*}{\textit{\makecell[l]{Sparse}}}
& \texttt{BM25} & Zero-shot & 0.182 & 0.279 & 0.164 & 0.134 & 0.109 & 0.163 & 0.161 & 0.170 \\
& \texttt{BM25 + RM3} & Zero-shot & 0.160 & 0.276 & 0.146 & 0.125 & 0.078 & 0.141 & 0.128 & 0.150 \\

\midrule
\multirow{3}{*}{\textit{\makecell[l]{Dense}}}
& \texttt{GritLM-7B} & SFT & 0.248 & 0.323 & 0.189 & 0.198 & 0.171 & 0.136 & 0.178 & 0.206 \\
& \texttt{GTE-Qwen-7B} & SFT & 0.306 & 0.364 & 0.178 & 0.246 & 0.132 & 0.222 & 0.148 & 0.228 \\
& \texttt{ReasonIR-8B} & SFT & 0.262 & 0.314 & 0.233 & 0.300 & 0.180 & 0.239 & 0.205 & 0.248 \\

\midrule
\multirow{3}{*}{\textit{\makecell[l]{Rerankers}}}
& \texttt{RankGPT4} & Zero-shot & 0.338 & 0.342 & 0.167 & 0.270 & 0.223 & 0.277 & 0.111 & 0.247 \\
& \texttt{RankZephyr-7B} & GPT4-distill & 0.219 & 0.237 & 0.144 & 0.103 & 0.076 & 0.137 & 0.166 & 0.155 \\
& \texttt{Rank-R1-14B} & GRPO (RL) & 0.312 & 0.385 & 0.212 & 0.264 & 0.226 & 0.189 & 0.275 & 0.266 \\

\midrule
\multirow{11}{*}{\textit{\makecell[l]{Query\\Expansions\\w/ \texttt{BM25}}}}
& \texttt{GenQR} & \multirow{11}{*}{Zero-shot} & 0.398 & 0.439 & 0.222 & 0.320 & 0.160 & 0.268 & 0.142 & 0.278 \\
& \texttt{GenQREnsemble} & & 0.424 & 0.471 & 0.231 & 0.320 & 0.156 & 0.266 & 0.172 & 0.291 \\
& \texttt{QA-Expand} & & 0.267 & 0.385 & 0.185 & 0.202 & 0.112 & 0.171 & 0.183 & 0.215 \\
& \texttt{Query2E} & & 0.239 & 0.376 & 0.181 & 0.187 & 0.121 & 0.190 & 0.162 & 0.208 \\
& \texttt{Query2Doc (ZS)} & & 0.255 & 0.380 & 0.172 & 0.200 & 0.126 & 0.192 & 0.184 & 0.216 \\
& \texttt{Query2Doc (FS)} & & 0.258 & 0.368 & 0.179 & 0.187 & 0.114 & 0.181 & 0.171 & 0.208 \\
& \texttt{Query2Doc (CoT)} & & 0.252 & 0.376 & 0.184 & 0.172 & 0.121 & 0.192 & 0.177 & 0.211 \\
& \texttt{LameR} & & 0.315 & 0.408 & 0.197 & 0.258 & 0.130 & 0.220 & 0.206 & 0.248 \\
& \texttt{MUGI} & & \underline{0.475} & \underline{0.502} & \underline{0.296} & 0.408 & \underline{0.189} & \underline{0.274} & 0.269 & 0.345 \\
& \texttt{CSQE} & & 0.320 & 0.455 & 0.209 & 0.265 & 0.137 & 0.215 & 0.201 & 0.257 \\
& \texttt{ThinkQE} & & 0.456 & \textbf{0.516} & 0.291 & \underline{0.420} & 0.186 & 0.268 & \underline{0.294} & \underline{0.347} \\[0.3em]

\cline{2-11}
\cline{2-11} \\[-0.87em]
& \textbf{\texttt{ADORE}} & Zero-shot & \textbf{0.501} & 0.490 & \textbf{0.332} & \textbf{0.468} & \textbf{0.232} & \textbf{0.279} & \textbf{0.347} & \textbf{0.379} \\

\bottomrule
\end{tabular}%
}
\caption{Results on BRIGHT benchmark (nDCG@10). All query reformulation baselines use GPT-4.1 as the backbone language model. Best scores in \textbf{bold}, second-best \underline{underlined}.}
\vspace{-0.5em}
\label{tab:bright_results}
\end{table*}

\paragraph{Datasets and evaluation.}
We evaluate \texttt{ADORE} on three benchmark families spanning complementary retrieval challenges. 
\textbf{Passage retrieval} includes the TREC DL tracks DL19~\citep{TREC2019}, DL20~\citep{TREC2020}, and DL-Hard~\citep{DL_HARD}, all built on MS MARCO passages~\citep{bajaj2018ms} with graded relevance judgments. 
\textbf{Cross-domain retrieval} uses five BEIR datasets~\citep{thakur2021beir}: SciFact, TREC-COVID, FiQA, DBPedia-Entity, and TREC-NEWS, covering scientific fact-checking, biomedical literature, financial opinion, entity retrieval, and news. 
\textbf{Reasoning-intensive retrieval} uses seven BRIGHT sub-domains~\citep{bright}: Biology, Earth Science, Economics, Psychology, Robotics, Stack Overflow, and Sustainable Living, where queries require reasoning beyond lexical or simple semantic matching. 
Following prior work~\citep{mugi,wang2023query2doc,csqe,thinkqe}, we report nDCG@10 as the primary metric. Dataset statistics are provided in Appendix~\ref{sec:dataset_stats}.

\paragraph{Language models.}
\texttt{ADORE} uses LLMs in two roles: reformulators to generate expansions and relevance assessors to evaluate retrieved documents. Our main results use GPT-4.1 for both roles. In ablation studies, we additionally evaluate DeepSeek-V3~\citep{deepseekv3} and Llama-3.3-70B~\citep{llama3} as alternative reformulators and assessors. Relevance assessment follows the UMBRELA prompting framework~\citep{upadhyay2024umbrela}, which elicits graded relevance judgments on a 0--3 scale.

\paragraph{Baselines.}
We compare \texttt{ADORE} against sparse retrieval and LLM-based query reformulation baselines. Sparse baselines include BM25 and BM25+RM3. Reformulation baselines include GenQR and GenQREnsemble~\citep{wang2023genqr,dhole2024genqrensemble}, Query2E~\citep{exploringresearch}, QA-Expand~\citep{qaexpand}, Query2Doc (CoT, FS, ZS)~\citep{wang2023query2doc}, LameR~\citep{lamer}, MUGI~\citep{mugi}, CSQE~\citep{csqe}, and ThinkQE~\citep{thinkqe}. Following prior reformulation studies \citep{querygym,bigdeli2026reproducibility}, all methods use BM25 as the first-stage retriever, with retrieval performed using the Pyserini toolkit~\citep{pyserini}. For fair comparison, all reformulation baselines are generated with GPT-4.1 using temperature 1.0 and a maximum length of 128 tokens, and are implemented with QueryGym~\citep{querygym}.

We also report results from dense retrievers, LLM-based retrievers, and rerankers as reference points rather than direct baselines, to contextualize how query expansion positions BM25 relative to learned retrieval paradigms. For TREC DL and BEIR, we include DPR~\citep{karpukhin2020dense}, ANCE~\citep{xiong2021approximate}, Contriever-FT~\citep{izacard2022unsupervised}, and BGE~\citep{xiao2024c}. For BRIGHT, we include stronger LLM-based retrievers and rerankers: GritLM-7B~\citep{gritlm}, GTE-Qwen-7B~\citep{gte_qwen}, ReasonIR-8B~\citep{reasonir}, RankGPT4~\citep{rankgpt}, RankZephyr-7B~\citep{rankzephyr}, and Rank-R1-14B~\citep{zhuang2025rank}.

\paragraph{Implementation details.}
\texttt{ADORE} runs for up to $T\!=\!5$ rounds, generating $n\!=\!5$ pseudo-passages per round and retrieving and assessing the top $K\!=\!10$ documents. The query construction function $f$ follows the repetition scheme of \citet{mugi}, where the original query is repeated and concatenated with the pseudo-passages, with the repetition weight calculated by the ratio of the appended expansion text length to the original query length. The relevance assessor decodes with temperature 0 and a maximum of 256 tokens. Adaptive termination triggers early stopping when either all top-10 documents receive the maximum relevance score (quality saturation) or the top-10 is unchanged for two consecutive rounds (coverage saturation, $\tau\!=\!1$).

\begin{table}[t]
\centering
\footnotesize
\newcommand{\redhlt}[1]{\textcolor{red!70!black}{#1}}
\newcommand{\greenhlt}[1]{\textcolor{green!45!black}{#1}}
\renewcommand{\arraystretch}{1.05}

\begin{tcolorbox}[
    colback=gray!3,
    colframe=gray!55,
    boxrule=0.6pt,
    arc=2pt,
    left=1pt,
    right=1pt,
    top=2pt,
    bottom=2pt,
    width=\columnwidth,
    enhanced
]


\textbf{\textit{Query:}} what type of tissue are bronchioles

\hfill
\textit{nDCG@10 = 0.441}

\vspace{0.45em}
\hrule
\vspace{0.55em}

\noindent
\begin{tabular*}{\linewidth}{@{}l@{\extracolsep{\fill}}r@{}}
\textbf{\textit{Zero-Shot Reformulation:}}
&
\textit{nDCG@10 = 0.610}
\end{tabular*}

\vspace{0.25em}

Bronchioles are small airways in the lungs that are
\redhlt{primarily composed of several types of tissue}.
Their walls consist of a lining of simple cuboidal or columnar epithelium,
surrounded by smooth muscle tissue. Unlike larger airways, bronchioles lack
\redhlt{cartilage} and \redhlt{mucous glands}. The smooth muscle tissue in
their walls allows bronchioles to regulate airflow by contracting or relaxing,
\redhlt{which is crucial for proper respiratory function}.

\vspace{0.55em}
\hrule
\vspace{0.55em}

\noindent
\begin{tabular*}{\linewidth}{@{}l@{\extracolsep{\fill}}r@{}}
\textbf{\textit{ADORE:}}
&
\textit{nDCG@10 = 0.916}
\end{tabular*}

\vspace{0.25em}

Bronchioles are distinguished by the composition of their walls, which lack
\greenhlt{hyaline cartilage} and are instead supported by smooth muscle and
\greenhlt{elastic fibers}. The epithelium lining the bronchioles
\greenhlt{transitions along their length}; it begins as
\greenhlt{ciliated columnar epithelium} in the larger bronchioles and shifts to
\greenhlt{simple ciliated cuboidal epithelium} in the smaller bronchioles.
The presence of smooth muscle tissue in the bronchiolar walls plays a critical
role in regulating airflow by constricting or relaxing as needed. As the
bronchioles decrease in size, \greenhlt{the epithelial cells also reduce in height}
and \greenhlt{the abundance of ciliated cells decreases}.

\end{tcolorbox}

\caption{Case study illustrating how iterative corpus-grounded feedback refines query reformulation. \redhlt{Red} marks vague phrases in the zero-shot query reformulation. \greenhlt{Green} marks domain-specific vocabulary acquired through iterative corpus-grounded feedback with \texttt{ADORE}.}
\vspace{-1em}
\label{tab:case_study}

\end{table}

\begin{figure*}[t]
    \centering
    \includegraphics[width=\textwidth]{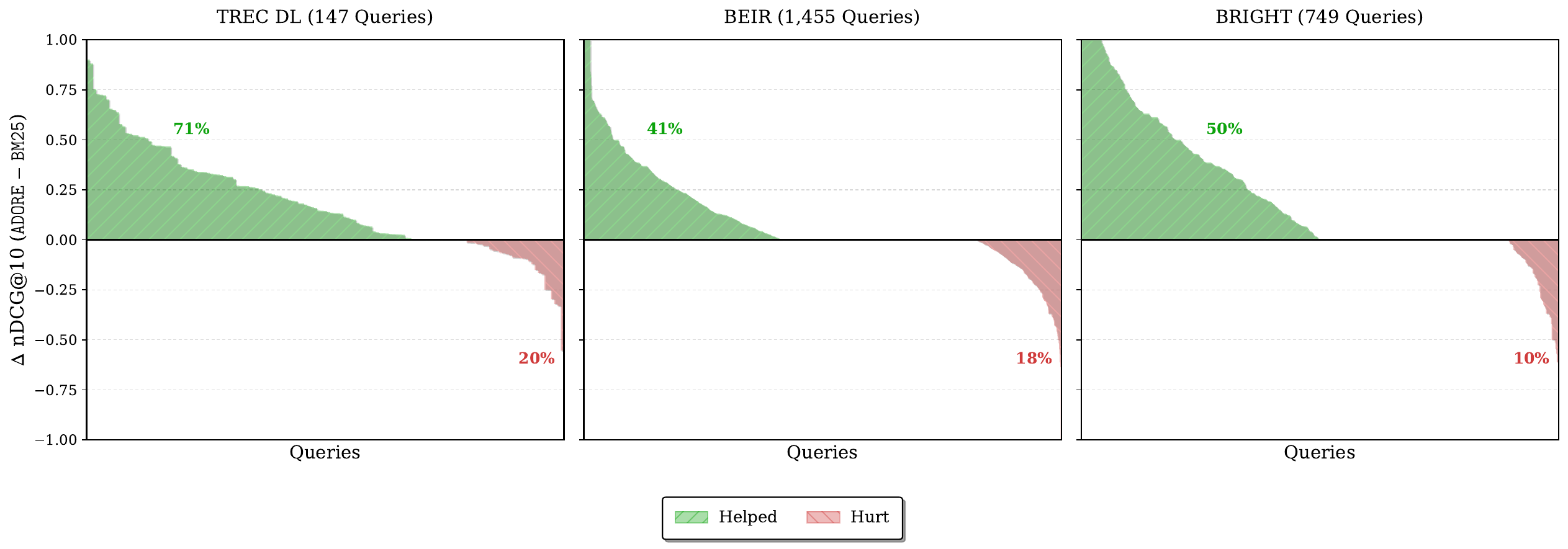}

    \caption{Per-query effect of \texttt{ADORE} relative to BM25 across benchmarks, measured by the change in nDCG@10. Green marks improved queries (Helped), where \texttt{ADORE} raises nDCG@10, and red marks degraded queries (Hurt), where it falls below. Percentages report the share of helped and hurt queries within each benchmark.}
    \label{fig:help_hurt}
\end{figure*}

\section{Main Results}
\label{sec:main_results}

Table~\ref{tab:comprehensive_results} and Table~\ref{tab:bright_results} report retrieval effectiveness on the TREC DL, BEIR, and BRIGHT benchmarks. All query expansion methods, including \texttt{ADORE}, use GPT-4.1 as the backbone LLM.
 
\paragraph{TREC Deep Learning and BEIR.}
Across these benchmarks, \texttt{ADORE} achieves the highest nDCG@10 among all query expansion methods on seven of eight evaluation settings. On TREC DL, \texttt{ADORE} consistently outperforms ThinkQE, the strongest iterative baseline. On DL20, for instance, \texttt{ADORE} improves over ThinkQE by 4.2\% relative. On the cross-domain BEIR tasks, notable gains emerge on TREC-COVID and FiQA, the latter reaching a 15.4\% relative improvement over ThinkQE, while the BEIR average rises to 0.554, a 3.6\% gain. Importantly, \texttt{ADORE} elevates BM25 to surpass BGE-base-en-v1.5, the strongest dense retriever, on all TREC tracks and reach near parity on the BEIR average, closing the gap between sparse and dense retrieval without any model training.

Table~\ref{tab:case_study} illustrates how \texttt{ADORE}'s feedback loop refines the reformulated query on a TREC DL 2020 example. The zero-shot pseudo-passage contains generic phrasing that fails to capture the corpus-specific terminology needed to retrieve highly relevant documents. After successive rounds of graded relevance feedback, \texttt{ADORE}'s pseudo-passage incorporates precise domain vocabulary from relevant documents, raising nDCG@10 to 0.916, up from 0.441 for the original query and 0.610 for the zero-shot reformulation.

\begin{table}[t]
\centering
\small
\setlength{\tabcolsep}{8pt}

\begin{tabular}{l cc}
\toprule
\textbf{Method} & \textbf{DL19} & \textbf{DL20} \\
\midrule
\texttt{BM25} & 0.506 & 0.480 \\
\texttt{ADORE} w/ lower-tier feedback only & 0.670 & 0.643 \\
\texttt{ADORE} w/ higher-tier feedback only & 0.696 & 0.709 \\
\textbf{\texttt{ADORE}} & \textbf{0.713} & \textbf{0.712} \\
\bottomrule
\end{tabular}

\caption{Ablation isolating the contribution of the positive and negative relevance tiers in \texttt{ADORE}'s feedback-conditioned prompt (nDCG@10).}
\vspace{-1em}
\label{tab:feedback_ablation}
\end{table}

\begin{figure*}[t]
    \centering

    \includegraphics[width=0.9\textwidth,
            trim=0cm 0.25cm 0cm 0cm,
        clip
        ]{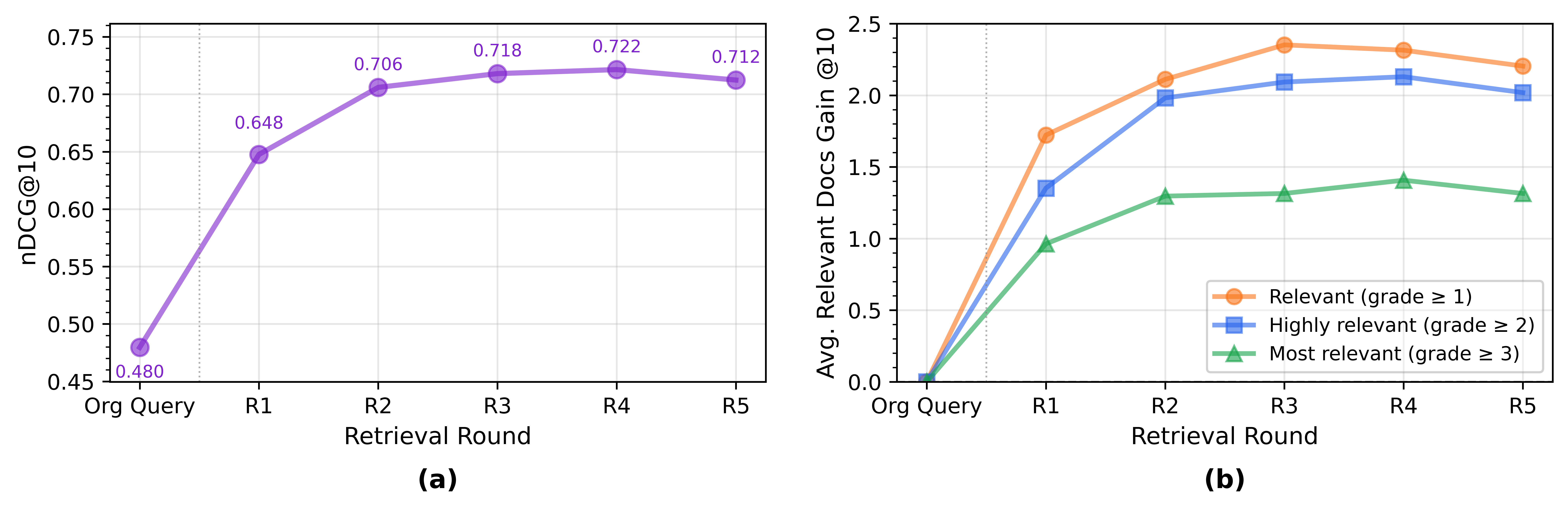}
    \caption{Per-round retrieval effectiveness on TREC DL 2020. (a) nDCG@10 at each round. (b) Average gain in relevant documents within the top-10 relative to the original query, stratified by TREC relevance grade.}
\vspace{-0.5em}
    \label{fig:convergence_analysis}
\end{figure*}

\paragraph{BRIGHT.}
On the reasoning-intensive BRIGHT benchmark, the benefits of iterative corpus-grounded reformulation are amplified. \texttt{ADORE} achieves an average nDCG@10 of 0.379, improving over ThinkQE by 9.2\% and over MUGI by 9.9\%, with the largest per-domain gains on Sustainable Living, where the margin over ThinkQE reaches 18.0\%, and Robotics at 24.7\%. These are domains where complex, multi-faceted queries demand precise lexical alignment that benefits from multiple rounds of feedback-driven refinement. \texttt{ADORE} ranks first on six of seven sub-domains; Earth Science is the sole exception, where ThinkQE leads by a small margin. The contrast with learned retrieval paradigms is especially striking: \texttt{ADORE} over BM25 surpasses all LLM-based dense retrievers, including the reasoning-specialized ReasonIR-8B, and all LLM-based rerankers, including Rank-R1-14B, by wide margins. These results indicate that on tasks requiring deep reasoning, iterative corpus calibration over a lexical retriever can substantially outperform both supervised dense encoders and LLM-based rerankers that operate without the feedback loop.


\paragraph{Per-Query Effectiveness Analysis.}
To characterize where \texttt{ADORE}'s gains originate, we conduct a per-query help-hurt analysis against BM25, classifying queries as helped, hurt, or unchanged based on their change in nDCG@10. Figure~\ref{fig:help_hurt} shows the distribution of per-query differences within each benchmark. Across all benchmarks, \texttt{ADORE} helps more queries than it hurts, indicating that improvements extend across a substantial proportion of the evaluated queries. The effect is most pronounced on TREC DL, where 71\% of queries improve while only 20\% degrade, and the same pattern holds on BEIR and BRIGHT, where helped queries outnumber hurt ones by a wide margin.

\begin{table}[t]
\centering
\small

\begin{tabular}{l ccc}
\toprule
\textbf{Relevance Assessor} & \textbf{DL19} & \textbf{DL20} & \textbf{DL-Hard} \\
\midrule
GPT-4.1              & 0.713 & 0.712 & 0.383 \\
DeepSeek-V3          & 0.720 & 0.698 & 0.389 \\
Llama-3.3-70B        & 0.732 & 0.691 & 0.391 \\
\midrule
\textbf{Reformulator} & \textbf{DL19} & \textbf{DL20} & \textbf{DL-Hard} \\
\midrule
GPT-4.1              & 0.713 & 0.712 & 0.383 \\
DeepSeek-V3          & 0.724	& 0.679	& 0.351           \\
Llama-3.3-70B        &0.684 &	0.695 &	0.367        \\
\bottomrule
\end{tabular}

\caption{Impact of swapping the relevance assessor (top) and reformulator (bottom) LLM on \texttt{ADORE} performance (nDCG@10). In each panel the other component is held fixed at GPT-4.1.}
\vspace{-1em}
\label{tab:llm_ablation}
\end{table}

\section{Ablation Studies}
\label{sec:ablation_results}

Here, we examine the core components and design decisions that drive ADORE's effectiveness.

\begin{table*}[t]
\centering
\small
\setlength{\tabcolsep}{5pt}

\resizebox{\textwidth}{!}{
\begin{tabular}{ll ccc ccccc c}
\toprule
\multirow{2}{*}{\textbf{Retriever}} & \multirow{2}{*}{\textbf{Query}} 
& \multicolumn{3}{c}{\textbf{TREC Deep Learning}} 
& \multicolumn{6}{c}{\textbf{BEIR Benchmark}} \\
\cmidrule(lr){3-5} \cmidrule(lr){6-11}
& & \textbf{DL19} & \textbf{DL20} & \textbf{DL-Hard} 
& \textbf{SciFact} & \textbf{COVID} & \textbf{FiQA} & \textbf{DBPed} & \textbf{NEWS} & \textbf{Avg.} \\
\midrule

\multirow{2}{*}{\texttt{BGE-base-en-v1.5}}
& Original Query & 0.702 & 0.677 & 0.379 & 0.741 & 0.780 & 0.407 & 0.407 & 0.442 & 0.555 \\
& +\texttt{ADORE} & \textbf{0.771} & \textbf{0.717} & \textbf{0.421} & \textbf{0.776} & \textbf{0.851} & \textbf{0.421} & \textbf{0.410} & \textbf{0.494} & \textbf{0.590} \\

\midrule
\multirow{2}{*}{\texttt{Contriever-FT}}
& Original Query & 0.621 & 0.632 & 0.396 & 0.677 & 0.596 & \textbf{0.329} & \textbf{0.413} & 0.428 & 0.489 \\
& +\texttt{ADORE} & \textbf{0.728} & \textbf{0.676} & \textbf{0.409} & \textbf{0.738} & \textbf{0.738} & 0.313 & 0.391 & \textbf{0.481} & \textbf{0.532} \\

\bottomrule
\end{tabular}
}
\caption{Dense retrieval effectiveness (nDCG@10) before and after \texttt{ADORE} reformulation. \emph{Original Query} uses the unmodified query, while \emph{+\texttt{ADORE}} uses the reformulated query produced by our framework.}
\vspace{-1em}
\label{tab:dense_reformulation}
\end{table*}

\subsection{Isolating Feedback Signals}
\label{sec:feedback_ablation}
 
\texttt{ADORE}'s feedback-conditioned prompt partitions retrieved documents into higher- and lower-tier relevance groups, reinforcing the lexical anchors of the higher-tier evidence while avoiding those of the lower-tier evidence. To isolate each tier's contribution, we compare the full model on DL19 and DL20 against higher-tier-only and lower-tier-only variants, with BM25 as the lower bound (Table~\ref{tab:feedback_ablation}).
 
Both tiers contribute but play distinct roles. Higher-tier feedback alone recovers most of \texttt{ADORE}'s effectiveness, reaching 0.696 and 0.709 on DL19 and DL20, making it the dominant driver of the gains. Lower-tier feedback alone is weaker, at 0.670 and 0.643, since suppressing false-positive vocabulary without higher-tier grounding gives the reformulator little to anchor on. Combining both is best, at 0.713 and 0.712, showing that the lower tier adds value on top of the higher one by directing the reformulator away from misleading terms.

\subsection{Effect of Iteration Depth}
\label{sec:convergence}
To understand how retrieval effectiveness evolves as \texttt{ADORE} progresses through successive feedback rounds, we track nDCG@10 and the number of relevant documents in the top-10 at each round on TREC DL 2020 (Figure~\ref{fig:convergence_analysis}). nDCG@10 rises sharply from 0.480 with the original query to 0.648 after R1 and 0.706 after R2, the first round to receive structured relevance feedback, before plateauing beyond R3. The gain in relevant documents follows the same trajectory (Figure~\ref{fig:convergence_analysis}(b)). These results show that \texttt{ADORE} converges rapidly, with the vast majority of improvement captured within three rounds, validating $T\!=\!5$ as a safe upper bound combined with adaptive early stopping. We extend these analyses in the appendix, reporting full per-round results in Appendix~\ref{sec:per_round}, termination behavior in Appendix~\ref{sec:termination}, and the average per-query cost in Appendix~\ref{sec:cost}.

\subsection{Impact of LLM Components}
\label{sec:llm_ablation}
 
To examine how sensitive \texttt{ADORE} is to the choice of its two LLM components, we conduct two controlled ablations on the TREC Deep Learning benchmarks (Table~\ref{tab:llm_ablation}). In the first, we fix GPT-4.1 as the reformulator and swap the relevance assessor across GPT-4.1, DeepSeek-V3, and Llama-3.3-70B. All three assessors yield competitive results with minimal variance, indicating that \texttt{ADORE} is agnostic to the assessor and applicable with open-weight alternatives. In the second, we fix GPT-4.1 as the assessor and vary the reformulator across the same three models. Performance again remains within a narrow range across benchmarks. Together, these results confirm that \texttt{ADORE}'s iterative feedback loop is robust to both component choices and does not depend on a single proprietary model.

\subsection{Retriever Generalizability}
\label{sec:dense_transfer}
 
Although \texttt{ADORE}'s reformulated queries are generated with BM25 as the retriever, we explore whether they can also improve dense retrieval by applying them to two widely adopted dense retrievers, BGE-base-en-v1.5 and Contriever-FT (Table~\ref{tab:dense_reformulation}). BGE improves on all eight settings with the BEIR average rising from 0.555 to 0.590, and Contriever-FT improves on six of eight, lifting its BEIR average from 0.489 to 0.532. These results confirm that \texttt{ADORE}'s corpus-calibrated expansions transfer effectively to dense retrieval without additional training. A full comparison of \texttt{ADORE} against all baselines over the BGE retriever is provided in Appendix~\ref{sec:bge_comparison} (Table~\ref{tab:bge_results}).

\section{Concluding Remarks}
 
We presented \texttt{ADORE}, an iterative query reformulation framework that couples expansion generation with explicit relevance assessment through a structured feedback loop. A dedicated assessor partitions retrieved documents into graded relevance tiers, enabling the reformulator to reinforce productive lexical signals and suppress retrieval drift across rounds. Experiments on TREC Deep Learning, BEIR, and BRIGHT show consistent improvements over strong baselines across nearly all evaluation settings. On the reasoning-intensive BRIGHT benchmark, \texttt{ADORE} elevates BM25 above both dense retrievers and LLM-based rerankers.

\section*{Limitations}
 
We acknowledge two limitations of our proposed method, \texttt{ADORE}. First, the iterative design introduces additional latency and cost over single-pass methods, as each round generates a new expansion from LLM-based assessment feedback, which may limit applicability in latency-sensitive settings despite adaptive termination. This is a cost-benefit trade-off, where the higher test-time cost yields notable effectiveness gains over single-pass reformulation. Second, since the LLMs' training data is not publicly available, we cannot rule out data contamination from benchmark queries or relevant documents seen during pretraining. This concern applies equally to \texttt{ADORE} and the other LLM-based query reformulation baselines we compare against.

\section*{Acknowledgments}
AI assistants were used for editing and proofreading the manuscript. All scientific claims, experimental design, analyses, and final content were verified by the authors.

\bibliography{custom}

\appendix
 \section*{Appendix}

\begin{figure*}[t]
\begin{tcolorbox}[colback=gray!3, colframe=black!60, boxrule=0.4pt, arc=2pt, left=6pt, right=6pt, top=6pt, bottom=6pt, fontupper=\small]
\textsc{System}\\[2pt]
{\small You are PassageGenGPT, an AI capable of generating concise, informative, and clear pseudo passages on specific topics.}\\[6pt]
\textsc{User}\\[2pt]
{\small Generate one passage that is relevant to the following query: `\texttt{\{query\}}'. The passage should be concise, informative, and clear}\\[6pt]
\textsc{Assistant}\\[2pt]
{\small Sure, here's a passage relevant to the query:}
\end{tcolorbox}

\caption{Zero-shot reformulation prompt.}
\label{fig:zeroshot_prompt}

\end{figure*}

\begin{figure*}[h!]
\begin{tcolorbox}[colback=gray!3, colframe=black!60, boxrule=0.4pt, arc=2pt, left=6pt, right=6pt, top=6pt, bottom=6pt, fontupper=\small]
\textsc{System}\\[2pt]
{\small You write corpus-grounded passages that expand a retrieval query toward more relevant documents. The retriever matches on exact lexical content, so reuse the distinctive lexical anchors of the relevant evidence verbatim --- synonyms or paraphrases will not match.}\\[0pt]

\textsc{User}\\[2pt]
{\small\textbf{Original Query}\quad \texttt{\{q0\}}}\\[0pt]

{\small\textbf{Relevant Evidence} {\normalfont\itshape (judged against the original query; the lexical signature of the relevant region of the corpus --- reuse its named entities, numbers, units, acronyms, and short technical phrasings verbatim)}}\\[3pt]
{\small\quad\colorbox{green!8}{\strut\,Score 3\,}\quad \texttt{\{score3\_block\}}}\\
{\small\quad\colorbox{yellow!15}{\strut\,Score 2\,}\quad \texttt{\{score2\_block\}}}\\[0pt]

{\small\textbf{Avoid Evidence} {\normalfont\itshape (retrieval false positives; their distinctive phrases are noise --- do NOT reuse them)}}\\[3pt]
{\small\quad\colorbox{orange!15}{\strut\,Score 1 --- intent-drifted\,}\quad \texttt{\{score1\_block\}}}\\
{\small\quad\colorbox{red!10}{\strut\,Score 0 --- irrelevant\,}\quad \texttt{\{score0\_block\}}}\\[0pt]

{\small\textbf{Instructions}}\\[3pt]
{\small Write one new passage that}
\begin{enumerate}[leftmargin=18pt, itemsep=1pt, topsep=2pt, label={\small\arabic*.}]
\item {\small Reuses distinctive lexical anchors from the relevant evidence verbatim (no synonym swaps for rare technical terms, named entities, units, numbers).}
\item {\small Targets a sub-aspect of the original query that the relevant evidence does not already saturate.}
\item {\small Reads like a real document drawn from the same corpus and register as the relevant evidence (no lists, no markdown, no meta-commentary).}
\item {\small Avoids the distinctive vocabulary of the avoid evidence.}
\item {\small Stays factually grounded in the relevant evidence; do not fabricate entities or statistics beyond what it supports.}
\end{enumerate}
\vspace{2pt}
{\small Return ONLY the passage text, nothing else.}
\end{tcolorbox}

\caption{Feedback-conditioned reformulation prompt.}
\label{fig:feedback_prompt}

\end{figure*}

\section{Dataset Statistics}
\label{sec:dataset_stats}

Details about the retrieval datasets are shown in Table~\ref{tab:dataset_stats}. The TREC Deep Learning tracks (DL19, DL20, DL-Hard) are built on the MS MARCO v1 passage collection, which contains approximately 8.8M passages. The BEIR datasets span diverse domains and vary considerably in corpus size, from 5,183 documents (SciFact) to over 4.6M (DBPedia-Entity). The BRIGHT benchmark~\citep{bright} consists of reasoning-intensive queries drawn from StackExchange communities, where relevant documents require complex understanding beyond lexical or simple semantic matching.
\begin{table}[t]
\centering

\scalebox{0.87}{
\begin{tabular}{lrr}
\toprule
\textbf{Dataset} & \textbf{\#Queries} & \textbf{\#Documents} \\
\midrule
\multicolumn{3}{c}{\textit{TREC Deep Learning}} \\
\midrule
DL19                      & 43    & \multirow{3}{*}{8,841,823} \\
DL20                      & 54    &  \\
DL-Hard (DLH)             & 50    &  \\
\midrule
\multicolumn{3}{c}{\textit{BEIR Benchmark}} \\
\midrule
SciFact                   & 300   & 5,183     \\
TREC-COVID (COVID)        & 50    & 171,332   \\
FiQA                      & 648   & 57,638    \\
DBPedia-Entity (DBPed)    & 400   & 4,635,922 \\
TREC-NEWS (NEWS)          & 57    & 594,977   \\
\midrule
\multicolumn{3}{c}{\textit{BRIGHT Benchmark}} \\
\midrule
Biology (Bio.)            & 103   & 57,359    \\
Earth Science (Earth.)    & 116   & 121,249   \\
Economics (Econ.)         & 103   & 50,220    \\
Psychology (Psy.)         & 101   & 52,835    \\
Robotics (Rob.)           & 101   & 61,961    \\
Stack Overflow (Stack.)   & 117   & 107,081   \\
Sustainable Living (Sus.) & 108   & 60,792    \\
\bottomrule
\end{tabular}}
\caption{Dataset statistics. Parenthetical labels denote the abbreviations used in the column headers of Tables~\ref{tab:comprehensive_results} and~\ref{tab:bright_results}.}

\label{tab:dataset_stats}
\end{table}

\section{Reformulation Prompts}
\label{sec:prompts}
 
This appendix provides the exact prompts used by the \texttt{ADORE} reformulator at each stage. Placeholder variables are shown in \texttt{monospace}.

\subsection{Zero-Shot Prompt (Round 1)}
\label{sec:prompt_zeroshot}

In the first round, no corpus feedback is available. The reformulator generates pseudo-passages from the query alone using the prompt shown in Figure~\ref{fig:zeroshot_prompt}.

\subsection{Feedback-Conditioned Prompt (Rounds $\geq 2$)}
\label{sec:prompt_iter}

In subsequent rounds, the reformulator receives the previous round's feedback pool partitioned into relevance tiers. The prompt instructs the model to reuse lexical anchors from relevant evidence, avoid vocabulary associated with irrelevant documents, and target under-represented aspects of the query. The full prompt is shown in Figure~\ref{fig:feedback_prompt}.

\section{Per-Round Retrieval Effectiveness}
\label{sec:per_round}

Table~\ref{tab:adore_rounds} extends the depth impact analysis in Section~\ref{sec:convergence} by reporting per-round nDCG@10 across all datasets. Across benchmarks, R2 delivers the largest single-round gain, and performance plateaus beyond R3. The reasoning-intensive BRIGHT datasets show larger gains from later rounds, suggesting complex queries benefit from additional corpus-grounded refinement.

\begin{table*}[h]
    \centering

    \scalebox{0.8}{%
    \begin{tabular}{l|l|ccccc}
    \toprule
    \textbf{Collection} & \textbf{Dataset} & \textbf{R1} & \textbf{R2} & \textbf{R3} & \textbf{R4} & \textbf{R5} \\
    \midrule
    \multirow{3}{*}{TREC DL} & DL19     & 0.687\scriptsize{(0.213)} & 0.711\scriptsize{(0.214)} & \textbf{0.719}\scriptsize{(0.223)} & 0.716\scriptsize{(0.225)} & 0.713\scriptsize{(0.223)} \\
    & DL20                              & 0.648\scriptsize{(0.231)} & 0.706\scriptsize{(0.223)} & 0.718\scriptsize{(0.218)} & \textbf{0.722}\scriptsize{(0.222)} & 0.712\scriptsize{(0.227)} \\
    & DL-Hard                           & 0.376\scriptsize{(0.321)} & 0.385\scriptsize{(0.310)} & \textbf{0.392}\scriptsize{(0.314)} & 0.386\scriptsize{(0.316)} & 0.383\scriptsize{(0.315)} \\
    \midrule
    \multirow{5}{*}{BEIR}    & SciFact  & 0.733\scriptsize{(0.347)} & 0.752\scriptsize{(0.354)} & 0.748\scriptsize{(0.352)} & \textbf{0.755}\scriptsize{(0.346)} & \textbf{0.755}\scriptsize{(0.345)} \\
    & COVID                             & 0.707\scriptsize{(0.238)} & 0.775\scriptsize{(0.240)} & \textbf{0.778}\scriptsize{(0.241)} & 0.772\scriptsize{(0.241)} & 0.774\scriptsize{(0.243)} \\
    & FiQA                              & 0.259\scriptsize{(0.320)} & 0.302\scriptsize{(0.350)} & 0.309\scriptsize{(0.355)} & 0.313\scriptsize{(0.353)} & \textbf{0.315}\scriptsize{(0.357)} \\
    & DBPedia                           & \textbf{0.414}\scriptsize{(0.266)} & 0.406\scriptsize{(0.271)} & 0.407\scriptsize{(0.270)} & 0.407\scriptsize{(0.276)} & 0.407\scriptsize{(0.274)} \\
    & NEWS                              & 0.508\scriptsize{(0.213)} & 0.528\scriptsize{(0.234)} & \textbf{0.531}\scriptsize{(0.234)} & 0.527\scriptsize{(0.231)} & 0.520\scriptsize{(0.228)} \\
    \midrule
    \multirow{7}{*}{BRIGHT}  & Biology  & 0.471\scriptsize{(0.355)} & 0.495\scriptsize{(0.384)} & 0.496\scriptsize{(0.388)} & 0.493\scriptsize{(0.386)} & \textbf{0.501}\scriptsize{(0.384)} \\
    & Earth Science                     & \textbf{0.521}\scriptsize{(0.322)} & 0.492\scriptsize{(0.339)} & 0.502\scriptsize{(0.347)} & 0.497\scriptsize{(0.342)} & 0.490\scriptsize{(0.342)} \\
    & Economics                         & 0.277\scriptsize{(0.325)} & 0.310\scriptsize{(0.360)} & 0.325\scriptsize{(0.375)} & 0.317\scriptsize{(0.381)} & \textbf{0.332}\scriptsize{(0.394)} \\
    & Psychology                        & 0.391\scriptsize{(0.362)} & 0.447\scriptsize{(0.383)} & 0.465\scriptsize{(0.381)} & 0.465\scriptsize{(0.384)} & \textbf{0.468}\scriptsize{(0.376)} \\
    & Robotics                          & 0.187\scriptsize{(0.260)} & 0.214\scriptsize{(0.301)} & 0.224\scriptsize{(0.315)} & 0.223\scriptsize{(0.311)} & \textbf{0.232}\scriptsize{(0.330)} \\
    & Stack Overflow                    & 0.271\scriptsize{(0.329)} & \textbf{0.281}\scriptsize{(0.335)} & 0.267\scriptsize{(0.334)} & \textbf{0.281}\scriptsize{(0.335)} & 0.279\scriptsize{(0.330)} \\
    & Sustainable Living                & 0.262\scriptsize{(0.302)} & 0.320\scriptsize{(0.352)} & 0.335\scriptsize{(0.369)} & 0.338\scriptsize{(0.369)} & \textbf{0.347}\scriptsize{(0.369)} \\
    \midrule
    \multicolumn{2}{l|}{\textbf{Average}} & 0.447 & 0.475 & 0.481 & 0.481 & 0.482 \\
    \bottomrule
    \end{tabular}%
    }
     \caption{Performance of \texttt{ADORE} across iterative rounds (nDCG@10). Each column represents the cumulative query expansion after $k$ rounds. Best score per dataset in \textbf{bold}. Standard deviation across queries shown in parentheses.}

    \label{tab:adore_rounds}
    \end{table*}

\begin{table*}[h]
\centering

\scalebox{0.8}{%
\begin{tabular}{l|ccc||ccccc|c}
\toprule
\multirow{2}{*}{\textbf{Method}} & \multicolumn{3}{c||}{\textbf{TREC Deep Learning}} & \multicolumn{6}{c}{\textbf{BEIR Benchmark}} \\
& \textbf{DL19} & \textbf{DL20} & \textbf{DLH} & \textbf{SciFact} & \textbf{COVID} & \textbf{FiQA} & \textbf{DBPed} & \textbf{NEWS} & \textbf{Avg.} \\
\midrule
\texttt{BGE-base-en-v1.5} & 0.702 & 0.677 & 0.379 & 0.741 & 0.780 & 0.407 & 0.407 & 0.442 & 0.555 \\
\quad \texttt{+ GenQR} & 0.702 & 0.690 & 0.387 & 0.748 & 0.778 & 0.392 & 0.356 & 0.464 & 0.548 \\
\quad \texttt{+ GenQREnsemble} & 0.703 & 0.683 & 0.357 & 0.759 & 0.800 & 0.403 & 0.376 & 0.475 & 0.562 \\
\quad \texttt{+ QA-Expand} & 0.737 & 0.707 & 0.374 & 0.737 & 0.795 & 0.416 & 0.401 & 0.470 & 0.564 \\
\quad \texttt{+ Query2E} & 0.697 & 0.642 & 0.378 & 0.742 & 0.774 & 0.392 & 0.325 & 0.445 & 0.536 \\
\quad \texttt{+ Query2Doc (ZS)} & 0.728 & \textbf{0.739} & 0.379 & \underline{0.761} & \underline{0.806} & 0.415 & \underline{0.431} & 0.476 & 0.578 \\
\quad \texttt{+ Query2Doc (FS)} & 0.727 & 0.714 & 0.407 & 0.752 & 0.804 & \underline{0.421} & 0.430 & 0.472 & 0.576 \\
\quad \texttt{+ Query2Doc (CoT)} & 0.713 & 0.672 & 0.376 & 0.758 & 0.798 & 0.401 & 0.368 & 0.433 & 0.552 \\
\quad \texttt{+ LameR} & 0.703 & 0.715 & 0.412 & 0.757 & 0.780 & 0.408 & 0.402 & 0.437 & 0.557 \\
\quad \texttt{+ MUGI} & 0.735 & \underline{0.720} & 0.404 & 0.757 & 0.802 & \textbf{0.429} & \textbf{0.440} & 0.490 & \underline{0.584} \\
\quad \texttt{+ CSQE} & \underline{0.755} & 0.714 & \underline{0.414} & 0.755 & 0.788 & 0.407 & 0.424 & 0.463 & 0.567 \\
\quad \texttt{+ ThinkQE} & 0.715 & 0.709 & 0.397 & 0.748 & 0.800 & 0.415 & 0.429 & \textbf{0.494} & 0.577 \\
\midrule
\quad \textbf{\texttt{+ ADORE}} & \textbf{0.771} & 0.717 & \textbf{0.421} & \textbf{0.776} & \textbf{0.851} & \underline{0.421} & 0.410 & \textbf{0.494} & \textbf{0.590} \\
\bottomrule
\end{tabular}%
}
\caption{Results on TREC Deep Learning and BEIR benchmarks using \texttt{BGE-base-en-v1.5} as the dense retriever (nDCG@10). All query reformulation baselines use GPT-4.1 as the backbone language model. Best scores in \textbf{bold}, second-best \underline{underlined}.}
\vspace{-0.5em}
\label{tab:bge_results}
\end{table*}

\section{Query Expansion Baselines over Dense Retrieval}
\label{sec:bge_comparison}

Table~\ref{tab:comprehensive_results} reports all query expansion baselines with BM25 as the retriever. To examine whether these methods generalize beyond sparse retrieval, we replicate the same comparison using BGE-base-en-v1.5 as the underlying dense retriever (Table~\ref{tab:bge_results}). The reformulated queries produced by each method are applied to BGE without any modification or retriever-specific tuning. \texttt{ADORE} achieves the highest average nDCG@10 across BEIR, outperforming all baselines including MUGI, while also obtaining the best scores on DL19, DL-Hard, SciFact, COVID, and NEWS. These results confirm that \texttt{ADORE}'s corpus-calibrated expansions remain effective when transferred to dense retrieval.

\begin{table}[t]
\centering
\small
\setlength{\tabcolsep}{6pt}

\scalebox{0.9}{
\begin{tabular}{l ccccc}
\toprule
\textbf{Benchmark} & \textbf{R1} & \textbf{R2} & \textbf{R3} & \textbf{R4} & \textbf{R5} \\
\midrule
TREC DL & 1.4\% & 14.3\% & 12.9\% & 23.8\% & 47.6\% \\
BEIR    & 0.1\% & \phantom{0}1.0\% & \phantom{0}2.1\% & 16.7\% & 80.1\% \\
BRIGHT  & 0.0\% & \phantom{0}0.1\% & \phantom{0}1.5\% & 16.4\% & 82.0\% \\
\bottomrule
\end{tabular}
}
\caption{Distribution of the ratio of queries by their termination round across benchmarks.}
\vspace{-1em}
\label{tab:termination_rounds}
\end{table}

\begin{table}[t]
\centering
\small
\setlength{\tabcolsep}{6pt}

\begin{tabular}{l ccc}
\toprule
\multirow{2}{*}{\textbf{Benchmark}} & \textbf{Quality} & \textbf{Coverage} & \textbf{Round} \\
 & \textbf{Saturation} & \textbf{Saturation} & \textbf{Budget} \\
\midrule
TREC DL & 29.3\% & 40.8\% & 29.9\% \\
BEIR    & \phantom{0}3.4\% & 37.8\% & 58.8\% \\
BRIGHT  & \phantom{0}0.1\% & 37.6\% & 62.2\% \\
\bottomrule
\end{tabular}
\caption{Distribution of the ratio of queries stopped by each of \texttt{ADORE}'s termination criteria across benchmarks.}
\vspace{-1em}
\label{tab:termination_reasons}
\end{table}

\section{Termination Behavior}
\label{sec:termination}

We analyze how \texttt{ADORE}'s adaptive termination behaves in practice by examining both the round at which each query stops and the criterion that triggers it, which together reveal how well the stopping rules match computation to query difficulty.

Table~\ref{tab:termination_rounds} reports the distribution of termination rounds. On TREC DL, queries stop across a wide spread of rounds, with only 47.6\% reaching the five-round budget. On the harder BEIR and BRIGHT queries, termination concentrates at the final round, with roughly 80\% and 82\% running to R5. Harder queries thus use the full budget, while easier ones converge sooner.

Table~\ref{tab:termination_reasons} reports the distribution of stopping criteria. On TREC DL, most queries stop early through quality or coverage saturation, and only 29.9\% use the full budget. On BEIR and BRIGHT, quality saturation rarely fires and roughly 59\% and 62\% instead run to the budget, consistent with their greater difficulty. Coverage saturation stays a steady contributor of around 38 to 41\% throughout, confirming that termination is adaptive rather than uniform.

\section{Cost-Benefit Analysis}
\label{sec:cost}
 
\begin{table}[t]
\centering
\small
\setlength{\tabcolsep}{5pt}

\resizebox{\columnwidth}{!}{%
\begin{tabular}{l r r r r r r}
\toprule
\textbf{Method} & \makecell{\textbf{Input}\\\textbf{tok/q}} & \makecell{\textbf{Output}\\\textbf{tok/q}} & \makecell{\textbf{API}\\\textbf{calls/q}} & \makecell{\textbf{Latency}\\\textbf{(s)}} & \makecell{\textbf{Cost}\\\textbf{(\$)}} & \textbf{nDCG@10} \\
\midrule
\texttt{ADORE} ($T\!=\!5$)   & 39,907 & 2,457 & 40.0 & 17.9 & 0.099 & \textbf{0.501} \\
\texttt{ADORE} ($T\!=\!3$)   & 23,700 & 1,348 & 30.2 & 11.8 & 0.058 & 0.496 \\
\texttt{ThinkQE} ($T\!=\!3$) & 6,149  & 767   & 6.0  & 10.0 & 0.018 & 0.456 \\
\texttt{MUGI}                & 898    & 632   & 5.0  & 13.0 & 0.007 & 0.475 \\
\texttt{CSQE}                & 2,750  & 254   & 2.0  & 3.0  & 0.008 & 0.320 \\
\texttt{Query2Doc (ZS)}      & 132    & 128   & 1.0  & 1.6  & 0.001 & 0.255 \\
\bottomrule
\end{tabular}%
}
\caption{Average per-query cost and effectiveness of \texttt{ADORE} and representative reformulation baselines on BRIGHT-Biology.}
\label{tab:cost_analysis}
\end{table}
Table~\ref{tab:cost_analysis} reports the average per-query cost and effectiveness of \texttt{ADORE} and four reformulation baselines on BRIGHT-Biology. Cost is estimated at GPT-4.1 pricing of \$2.00 and \$8.00 per 1M input and output tokens, respectively. As shown, retrieval effectiveness does not consistently increase with computational cost. MUGI outperforms ThinkQE while requiring fewer tokens and API calls, and also surpasses CSQE at a lower cost. These comparisons suggest that the reformulation strategy, rather than computational expenditure alone, determines how effectively additional computation translates into retrieval gains. Within \texttt{ADORE}, the final two rounds yield modest additional gains, so reducing the round budget from five to three retains nearly all of its effectiveness at roughly 58\% of the five-round cost, an efficient operating point when compute is constrained. Its lead over the cheaper baselines nonetheless far exceeds the gap between its three- and five-round settings, indicating that a reduced round budget preserves a substantial effectiveness advantage. In high-stakes, reasoning-intensive domains such as biomedical and legal, where surfacing the correct evidence is critical, the objective is maximal effectiveness, which \texttt{ADORE} attains with a bounded overhead that is small against the cost of retrieving the wrong evidence.

\end{document}